\documentstyle[aps,prl,multicol,epsf]{revtex}

\begin{document}

\draft

\newcommand \be {\begin{equation}}
\newcommand \ee {\end{equation}}
\newcommand \eps {\epsilon}
\newcommand \bi {\bibitem}
\newcommand \s {\sigma}
\newcommand \lan {\langle}
\newcommand \ran {\rangle}
\newcommand \dd {{\mathrm{d}}}
\newcommand \qea {\mbox{$q_{\scriptscriptstyle {\rm EA}}$}}

\title{Absence of aging in the remanent magnetization in
Migdal-Kadanoff spin glasses}

\author{F. Ricci-Tersenghi$^a$ and F. Ritort$^b$}
\address{
$^a$ Abdus Salam International Center for Theoretical Physics,
Condensed Matter Group\\
Strada Costiera 11, P.O. Box 586, 34100 Trieste (Italy)\\
$^b$ Department of Physics, Faculty of Physics, University of
Barcelona\\
Diagonal 647, 08028 Barcelona (Spain)
}

\date{October 23, 1999}
\maketitle

\begin{abstract}
We study the non-equilibrium behavior of three-dimensional spin
glasses in the Migdal-Kadanoff approximation, that is on a
hierarchical lattice.  In this approximation the model has an unique
ground state and equilibrium properties correctly described by the
droplet model.  Extensive numerical simulations show that this model
lacks aging in the remanent magnetization as well as a maximum in the
magnetic viscosity in disagreement with experiments as well as with
numerical studies of the Edwards-Anderson model.  This result strongly
limits the validity of the droplet model (at least in its simplest
form) as a good model for real spin glasses.
\end{abstract} 

\pacs{PACS numbers: 75.10.Nr, 05.50.+q, 75.40.Gb, 75.40.Mg}

\begin{multicols}{2}
\narrowtext

Spin glasses are disordered magnets which for low impurity
concentrations above the Kondo regime display interesting
non-equilibrium phenomena.  In particular, a freezing of the dynamics
appears at a temperature $T_c$ below which slow relaxation phenomena
manifest through non-stationary effects in the zero-field-cooled
magnetization.  In this regime different non-equilibrium phenomena
have been observed such as aging, remanence and several memory as well
as chaotic effects~\cite{EXPER}.  Despite the great activity devoted
to understand the nature of the low temperature phase in
three-dimensional spin glasses (numerical simulations, experiments and
theory, see the review~\cite{YOUNG_BOOK}) still many questions
regarding the ground state (e.g.\ its shape and its uniqueness) and
the type of excitations remain unanswered.

The mean-field picture for spin glasses~\cite{MPV} (i.e.\ the results
obtained for the Sherrington-Kirkpatrick model), despite its great
theoretical interest, it is not able to furnish a real space picture of
the type of excitations present in spin glasses. To fill this gap, and
based on domain wall scaling arguments (initially proposed by McMillan
and Bray and Moore), Fisher and Huse proposed what has been termed
as {\em droplet model} for spin glasses~\cite{DROPLET}. In the droplet
model there are two unique ground states related by spin inversion
symmetry. Thermal fluctuations activate droplets which are supposed to
be compact domains of typical size $L$ and fractal surface of
dimension $d_s\ge d-1$. These excitations cost a free energy which
grows like $\Upsilon(T)L^{\theta}$ where $\theta$ is a
zero-temperature exponent and $\Upsilon(T)$ is a temperature dependent
stiffness constant. The idea that excitations in spin glasses are
compact droplets is the simplest description that finds its most
successful application in the study of phase transitions in ordered
systems. Despite its inherent simplicity, the droplet model has a
severe limitation, i.e.\ its main assumptions remain to be proven from
a correct microscopic theory.  If one of its key assumptions were
wrong then the whole set of predictions coming out from the model
should need to be revisited.

Now, what is an appropriate microscopic model which describes the
spin glass transition? The simplest proposal was putted forward by
Edwards and Anderson almost twenty years ago~\cite{EA}, who introduced
a random bond nearest-neighbor interaction model, the so called
Edwards-Anderson model (EA model). It is widely believed that the EA
model is a real spin glass, i.e.\ it reproduces the major part of
results experimentally measured in the laboratory. So the question is
whether the droplet model \cite{DROPLET} is the appropriate theory to
describe the phenomenology already contained in the EA model. Despite
of the large number of numerical works devoted to this question (see
the reviews~\cite{RIEGER,REVIEW}) there is still no an universal
agreement on it.

Our work has been motivated by recent results by the Manchester
group~\cite{MANCHESTER} who found that finite-size effects in the
Migdal-Kadanoff approximation (MKA) of the three-dimensional EA model
are mean-field like.  While in the thermodynamical limit the MKA is
known to be described by the droplet model with $d_s=d-1$ and
$\theta\simeq 0.26$~\cite{SY,BM2}.  So the Manchester group suggested
that the droplet model could also explain the vast majority of
numerical simulation results for the EA model obtained during the last
decade (which on the other hand, have been taken by the Rome group as
evidences against the droplet picture). This is an
interesting observation whose physical meaning and consequences need
to be better understood and was already anticipated quite long ago in
a theoretical study of the one-dimensional spin-glass
chain~\cite{BM1}.  This controversy has been centered around the study
of the spin-glass equilibrium properties. So, it is now time to check
whether non-equilibrium behavior is well reproduced by the droplet
model. This is of the outmost importance because experimental
measurements in spin glasses in the low temperature regime are always
taken in the out-of-equilibrium regime.

In this paper we want to show that the droplet model lacks one of the
key features of real spin glasses found in the laboratory, i.e.\ the
aging in the zero-field-cooled magnetization. Consequently, the
physics contained in the droplet model corresponds to a limited class
of disordered systems being far from what is observed in real spin
glasses.

\begin{figure}
\epsfxsize=0.6\columnwidth
\vspace{-0.4cm}
\hspace{0.15\columnwidth}
\epsffile{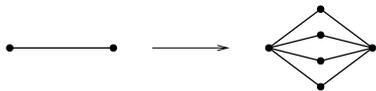}
\caption{Elementary step in the construction of the hierarchical
lattice, where the MK approximation is exact.}
\label{FIG1}
\end{figure}

The EA model in the presence of a field is defined by, 
\begin{equation}
{\cal H} = -\sum_{(i,j)}\,J_{ij}\s_i\s_j -h\sum_i\s_i
\label{eq1}
\end{equation}
where the site indexes run on the nodes of a cubic lattice, $(i,j)$
stands for nearest neighbor pairs, the spins take values $\s_i=\pm 1$
and the couplings are extracted from a Gaussian distribution of zero
average and unit variance.  Following~\cite{MANCHESTER} we will
consider the three dimensional EA model in the MKA which amounts to
consider a hierarchical lattice that is constructed iteratively by
replacing each bond by eight bonds as indicated in
fig.~\ref{FIG1}. Denoting by $g$ the number of generations then the
total number of bonds is $8^g$ which corresponds to the number of
sites for a cubic lattice with lattice size $L=2^g$.

The order parameter can be defined through the equilibrium
autocorrelation function,

\begin{equation}
\qea = \lim_{t\to\infty} \lim_{V\to\infty}
\frac{\sum_{i=1}^V x_i\langle\s_i(t)\s_i(0)\rangle}{\sum_{i=1}^V x_i}
\quad ,
\label{eq2}
\end{equation}
where $V=8^g=L^3$ is the volume and the averages $\langle..\rangle$
are taken over dynamical histories starting from different equilibrium
initial conditions at time 0. The parameters $x_i$ are
weights which may consider the fact that a given site is connected
with a different number of bonds depending on its generation level
(i.e.\ depending on which iteration in the recursive construction of
the lattice that site was generated).  Our results concentrate on the
choice $x_i=1$, i.e.\ all sites are identically weighted.  However
also the results obtained with $x_i=c_i$, where $c_i$ is the
connectivity of site $i$ (so all bonds are identically weighted)
corroborate our conclusions~\cite{FOOTNOTE1}.

We have concentrated our attention in the study of the relaxational
dynamics in the low-temperature phase $T<T_c \simeq 0.88$~\cite{SY}
below which \qea\ defined in Eq.(\ref{eq2}) is different from zero. We
have used Monte Carlo dynamics with Metropolis algorithm and random
updating~\cite{FOOTNOTE2}.  In our runs we follow the typical aging
experiment scheduling, that is: at $t=0$ we quench the system from
infinite temperature to a finite one $T<T_c$ without magnetic field,
letting the system evolve for a time $t_w$.  At time $t_w$ we switch
the magnetic field on.  For subsequent times ($t>t_w$) the system
continues to relax in a magnetic field $h$ and then we measure the
following two quantities: a) the autocorrelation function
\begin{equation}
C(t,t_w) = \frac{\sum_{i=1}^V x_i \s_i(t) \s_i(t_w)}{\sum_{i=1}^V x_i}
\quad ,
\label{eq3}
\end{equation}
and b) the zero-field-cooled susceptibility defined by
\begin{equation}
\chi_{ZFC}(t,t_w) = \lim_{h \to 0} \frac{M_{ZFC}(t,t_w)}{h} \quad ,
\label{eq4}
\end{equation}
where
\begin{equation}
M_{ZFC}(t,t_w) = \frac{\sum_{i=1}^V x_i [\s_i(t) - \s_i(t_w)]}
{\sum_{i=1}^V x_i} \quad .
\label{eq5}
\end{equation}
The limit in eq.(\ref{eq4}) is usually ignored, because we always work
in the linear response regime.  All the data we present have been
obtained with a magnetic field of intensity $h=0.1$ and we have
checked that the same susceptibility is obtained by doubling the
perturbing field.

We have performed extensive numerical simulations for $g\!=\!5$
($L\!=\!32$) and $g\!=\!6$ ($L\!=\!64$) at two different temperatures
($T=0.7,0.5$) and for many values of $t_w$.  We obtain the same
results for both temperatures. Here we present only those for $T=0.7$,
while the complete set of data will be reported elsewhere.  Note that
the ratio $T/T_c \simeq 0.8$ used here is very similar to that used in
many experiments~\cite{SITGES}.

\begin{figure}
\epsfxsize=0.95\columnwidth
\epsffile{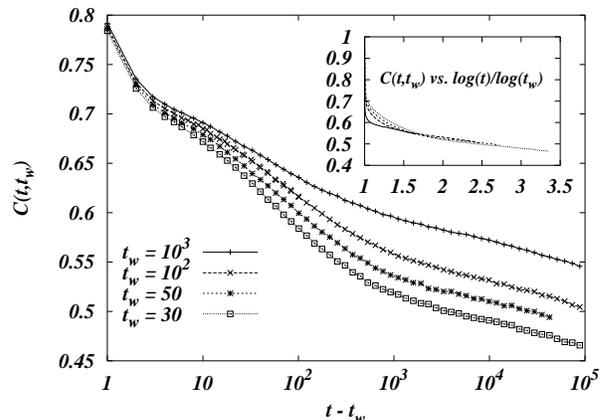}
\caption{The correlation functions for $g=6$, $T=0.7$ and different
$t_w$ clearly show aging. In the inset the same curves as a function
of the scaling variable $\log(t)/\log(t_w)$.}
\label{FIG2}
\end{figure}

In fig.~\ref{FIG2} we show the autocorrelation function for different
values of $t_w$. One can observe the presence of aging characteristic
of many glassy systems.

Following general assumptions, in the asymptotic regime $t_w\to\infty$,
the correlation function decomposes in two terms each one governing a
different time regime. In the quasi-equilibrium regime $t-t_w<<t_w$ the
system is in some sort of local equilibrium and correlation functions
are time translational invariant. In the aging regime $t-t_w>>t_w$ the
system ages and correlation functions depend on both times through non
trivial scaling relations. So in general one can write \cite{BCKM},
\begin{equation}
C(t,t_w) = C_{st}(t-t_w) + C_{aging}(t,t_w)
\label{eq6}
\end{equation}
with $\lim_{\tau \to \infty} C_{st}(\tau) = \qea$. In equilibrium the
aging part vanishes and one recovers the previous result of
eq.(\ref{eq2}).

The very difference between the experimental data and the EA model on
one hand and the MKA and the droplet model on the other is the large
times scaling of dynamical functions.  As can be seen in the inset of
fig.~\ref{FIG2}, in the MKA we find that the aging part of the
autocorrelation function is well described, in the large times limit,
by a function of the ratio $\log(t)/\log(t_w)$.  On the other hand in
experiments and in the EA model the scaling is far from the
$\log(t)/\log(t_w)$ and similar to $t/t_w$.  Because of the use of the
scaling variable, in the inset of fig.~\ref{FIG2} the data
corresponding to the quasi-equilibrium regime collapse on the line
$\log(t)/\log(t_w)=1$. Thus we can estimate the value of \qea\ as the
limit of the scaling function for $\log(t)/\log(t_w) \to 1^+$, i.e.\
$\qea \simeq 0.6$ (a value compatible with~\cite{MANCHESTER}).

\begin{figure}
\epsfxsize=0.95\columnwidth
\epsffile{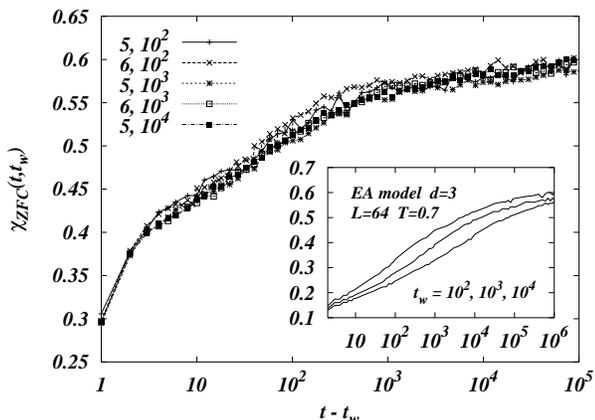}
\caption{The zero-field-cooled susceptibility data do not show any
aging. In the legend we report the values of $g$ and $t_w$. In the
inset we show the same quantity measured in the EA model, which
resembles a lot the experimental data.}
\label{FIG3}
\end{figure}

Our most striking result is found for the zero-field-cooled
susceptibility $\chi_{ZFC}(t,t_w)$ shown in fig.~\ref{FIG3}. In the
MKA there is no dependence of the susceptibility on $t_w$.  We believe
that such a result, which is characteristic of droplet models and
kinetic growth, makes the droplet model, at least in its simplest
form, inadeguate for the description of the EA one.  Aging in both
zero-field-cooled and field-cooled magnetization is so commonly found
in experiments on spin glasses that it is not clear to us how this
result can be explained by the standard droplet theory.  Note also
that the peak in the magnetic viscosity $S(t,t_w) = \partial
\chi_{ZFC}(t,t_w) / \partial \log(t-t_w)$ (experimentally very well
observed~\cite{UPSALA}) is completely absent in the MKA.  We should
remind that aging in $\chi_{ZFC}(t,t_w)$, with a peak in the
$S(t,t_w)$, is naturally found in the EA model (see inset of
fig.~\ref{FIG3}) as well as in mean-field models.  Then it remains to
be explained why these aging effects are naturally and easily observed
in the EA model and not in the MKA.

\begin{figure}
\epsfxsize=0.95\columnwidth
\epsffile{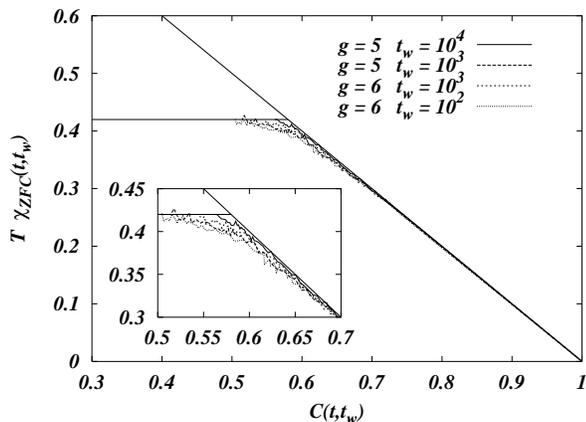}
\caption{The absence of response in the aging regime is suggested by
the rapid convergence of the $\chi_{ZFC}$ versus $C$ curves to the
plotted lines. Note that the horizontal line is an upper bound for the
data.}
\label{FIG4}
\end{figure}

Finally we consider the analysis of the fluctuation-dissipation ratio
useful to compare the results obtained in the MKA with those obtained
in the EA and coarsening models~\cite{FDT}. In the quasi-equilibrium
regime ($t-t_w<<t_w$) the system is in local equilibrium. Consequently
both correlation and susceptibility are time-traslational invariant
and the fluctuation-dissipation theorem (FDT) is satisfied,
\begin{equation}
T \chi_{ZFC}(t-t_w) = 1 - C(t-t_w) \quad .
\label{eq7}
\end{equation}
In the aging regime ($t-t_w>> t_w$) the system ages and FDT is
violated. Then it is useful to define the so called
fluctuation-dissipation ratio~\cite{FRARIE}
\begin{equation}
X(t,t_w) = \frac{T R(t,t_w)}{\frac{\partial C(t,t_w)}{\partial t_w}}
\quad ,
\label{eq8}
\end{equation}
which, in the asymptotic long-times limit $t,t_w \to \infty$, may be
uniquely expressed as function of the correlation $C(t,t_w)$ yielding
\begin{equation}
T \chi_{ZFC}(t,t_w) = \int_{C(t,t_w)}^1 X(C)\,\dd C \quad .
\label{eq9}
\end{equation}
Moreover the $X(C)$ can be related to equilibrium
quantities~\cite{FRAMEPAPE}.  The previous expression reduces to
Eq.(\ref{eq7}) in the quasi-equilibrium regime where $X\!=\!1$. A plot
of $T \chi_{ZFC}(t,t_w)$ as a function of $C(t,t_w)$ is expected to
show two different behaviors. For $\qea\!<\!C\!<\!1$ we have $X\!=\!1$
and so the curve $T \chi_{ZFC}$ versus $C$ has slope $-1$.  For
$C\!<\!\qea$ the $X$ may be a non-vanishing function of $C$ and we
have $T \chi_{ZFC}(t,t_w) = (1-\qea) + \int_{C(t,t_w)}^{\qea} X(C) \dd
C$. In coarsening models, $X\!=\!0$ for $C\!<\!\qea$ and so the
function $\chi_{ZFC}(C)$ is flat for $C\!<\!\qea$.  In fig.~\ref{FIG4}
we show the $\chi_{ZFC}$ as a function of $C$ for different values of
$g$ and $t_w$, which show that the behavior rapidly converges to that
of coarsening models and strongly differs from that observed in
finite-dimensional EA spin glasses~\cite{FDT}.  The horizontal line in
fig.~\ref{FIG4} is the infinite time limit of the susceptibility,
extrapolated from the data of fig.~\ref{FIG3} and from those for the
field-cooled magnetization (not shown).  It is an upper bound for the
plotted curves, thanks to the positiveness of $X$ ratio.  From
fig.~\ref{FIG4} we can also get an estimate for the \qea\ order
parameter, defined as the abscissa value where the curves leave the
FDT line ($T \chi_{ZFC}=1-C$). Very reasonably this point is
converging, in the large times limit, near to the intersection of the
two lines, giving $\qea \simeq 0.6$ (as already found from the data of
fig.~\ref{FIG2}).

Fig.~\ref{FIG4} adds more evidence on the fact that spin
glasses in the MKA do not capture all the key features of
finite-dimensional spin glasses as we know them from the
three-dimensional EA model.

To summarize, we have shown that in the MKA spin glasses do not show
aging in the integrated response function. This aging is
experimentally observed in real spin glasses through zero-field-cooled
and field-cooled measurements being one of the key features which
distinguishes spin glasses from other disordered systems. The study of
the fluctuation-dissipation ratio suggests that relaxation in this
model is driven by coarsening, like in conventional ferromagnets.  One
could argue that these results for the MKA are not extensible to the
droplet model because, in the general case, the inequality $d_s \ge
d-1$ could restore aging.  Despite this possibility our results
unambiguously show that the MK model is not a good model for realistic
spin glasses. A new class of excitations or droplets must be present
in spin glasses. The droplet model in its simplest version does not
capture the physics behind real spin glasses.

One possible generalization of the droplet model (which would be no
longer simply droplet) is to consider two kind of basic excitations in
a spin glass: on small length scale the usual droplets and, in
addition, system-size scale collective
rearrangements~\cite{MARTIN}. The second kind of excitation are, at
present, ignored in the droplet model (they are exponentially rare),
but they could be responsible for the many mean-field like features
observed in finite-dimensional spin glasses.  In terms of a very
simplified energy landscape the two excitations would correspond,
respectively, to the local movements of the system in a single
``valley'' and to the jumps from one valley to an other one.  In our
opinion a new theory comprehensive of the small scale droplets and the
system-size scale excitations (with a clear real space picture) would
be welcome and could hopefully terminate the longstanding discussion
on finite-dimensional spin glasses.

{\bf Acknowledgments}.  We thank J.P.~Bouchaud and S.~Franz for
discussions. F.R. is supported by the Ministerio de Educaci\'on y
Ciencia in Spain (PB97-0971).

\hspace{-2cm}

\end{multicols}
\end{document}